\documentclass[journal=aamick,manuscript=article,layout=onecolumn]{achemso}
\usepackage{hyperref}
\usepackage{mciteplus}
\usepackage{graphicx}
\usepackage{epstopdf} 
\usepackage{dcolumn}
\usepackage{bm}
\usepackage[cmex10]{amsmath}
\usepackage[utf8]{inputenc}
\usepackage{float}
\usepackage{amssymb}
\usepackage[T1]{fontenc}
\usepackage{natbib}
 \usepackage{booktabs}
 \usepackage{multirow}
\usepackage{epsfig}

\usepackage[T1]{fontenc} 

\author{Piotr Ogrodnik}
\email{piotr.ogrodnik@pw.edu.pl}
\affiliation{AGH University of Science and Technology, Institute of Electronics, 30-059 Kraków, Poland}
\altaffiliation{Warsaw University of Technology, Faculty of Physics, 00-662 Warsaw, Poland}

\author{Krzysztof Grochot}
\affiliation{AGH University of Science and Technology, Institute of Electronics, 30-059 Kraków, Poland}
\altaffiliation{AGH University of Science and Technology, Faculty of Physics and Applied Computer Science, 30-059 Kraków, Poland}

\author{Łukasz Karwacki}
\affiliation{Institute for Theoretical Physics, Utrecht University, Princetonplein 5, 3584 CC Utrecht, Netherlands}
\altaffiliation{Institute of Molecular Physics, Polish Academy of Sciences, ul. M.Smoluchowskiego 17, 60-179 Poznań, Poland }

\author{Jarosław Kanak}
\affiliation{AGH University of Science and Technology, Institute of Electronics, 30-059 Kraków, Poland}

\author{Michał Prokop}
\affiliation{Catalan Institute of Nanoscience and Nanotechnology (ICN2), CSIC and BIST, Campus UAB, Bellaterra, 08193 Barcelona, Spain}

\author{Jakub Chęciński}
\affiliation{AGH University of Science and Technology, Institute of Electronics, 30-059 Kraków, Poland}

\author{Witold Skowroński}
\affiliation{AGH University of Science and Technology, Institute of Electronics, 30-059 Kraków, Poland}

\author{Sławomir Ziętek}
\affiliation{AGH University of Science and Technology, Institute of Electronics, 30-059 Kraków, Poland}

\author{Tomasz Stobiecki}
\affiliation{AGH University of Science and Technology, Institute of Electronics, 30-059 Kraków, Poland}
\altaffiliation{AGH University of Science and Technology, Faculty of Physics and Applied Computer Science, 30-059 Kraków, Poland}

\date{\today}

\title{Study of Spin-Orbit Interactions and Interlayer Ferromagnetic  Coupling in Co/Pt/Co Trilayers in Wide Range of Heavy Metal Thickness}

\abbreviations{FMR, SOT-FMR, SOT, DW}
\keywords{Ferromagnetic Resonance, Spin Hall Effect, Magnetoresistance, Spin Orbit Torques, Spin Accumulation, Edelstein-Rashba Effect, X-Ray Diffraction}

\begin{document}




\begin{abstract} 
The spin-orbit torque, a torque induced by a charge current flowing through the heavy-metal conducting layer with strong spin-orbit interactions, provides an efficient way to control the magnetization direction in heavy-metal/ferromagnet
nanostructures, required for applications in the emergent magnetic technologies like random access memories, high-frequency nano oscillators, or bio-inspired neuromorphic computations. We study the interface properties, magnetization dynamics, magnetostatic features and spin-orbit interactions within the multilayer system Ti(2)/Co(1)/Pt(0-4)/Co(1)/MgO(2)/Ti(2) (thicknesses in nanometers) patterned by optical lithography on micrometer-sized bars. In the investigated devices, Pt is used as a source of the spin current and as a non-magnetic spacer with variable thickness, which enables the magnitude of the interlayer ferromagnetic exchange coupling to be effectively tuned. We also find the Pt thickness-dependent changes in magnetic anisotropies, magnetoresistance, effective Hall angle and, eventually, spin-orbit torque fields at interfaces. The experimental findings are supported by the relevant interface structure-related simulations,  micromagnetic, macrospin, as well as the spin drift-diffusion models.  Finally, the contribution of the spin-orbital Edelstein-Rashba interfacial fields is also briefly discussed in the analysis.

\end{abstract}


\maketitle

\section{Introduction}
The magnetic multilayer structures consisting of thin ferromagnetic (F) layers and non-magnetic spacers are known to exhibit a plenty of phenomena, among which one can find those extensively studied for the last decades  like anisotropic, giant and tunneling magnetoresistace or spin-transfer torque effect (STT)\cite{ralphstt,brataas2012} and recently current-driven spin orbit torque (SOT) magnetization switching.\cite{manchon2019} These effects are widely exploited  in the spintronic devices - magnetic random access memories (MRAM), like STT-MRAM and SOT-MRAM,\cite{zhang2021field,bhatti2017spintronics,ikegawa2020magnetoresistive} as well as may be exploited in magnetic sensors (including magnetic nanoparticles) and nanooscillators.\cite{dieny2020,hirohata2020review} Such devices include non-magnetic layers that are crucial for their features and performance. These layers may be both insulating (e.g. MgO in magnetic tunnel junctions) or metallic (e.g. Cu, Au in GMR devices).\cite{hirohata2020review} Recently,the non-magnetic layers made of heavy metallic (HM) elements (W, Ta, Pt and their alloys\cite{obstbaum16,meinert,wang2017spin}) are extensively studied because of their large spin-orbit coupling (SOC).\cite{song2020spin} Such layers combined with ferromagnetic ones (typically Co, CoFeB) are expected to have new spin transport properties related to the SOC, e.g. spin Hall effect (SHE) and Rashba-Edelstein effect (REE).\cite{sinova2015spin,edelstein1990spin} Although the SHE occurs in single HM layer,\cite{hirsh1999} it is detectable in heterostructures with ferromagnets only, such as F/HM bi- \cite{mihajlovic2016pt} and F/HM/F trilayers,\cite{lim2020effect,staszek2019}. In these structures the spin-polarized electrons can accumulate at the HM/F interfaces and then may be efficiently injected into F layer exerting the spin-orbit torque (SOT) on its magnetic moment. This effect has been predicted theoretically\cite{chentheory,choitheory,hals2013phenomenology} and reported in many experimental works on SOT-induced magnetic dynamics\cite{ralphsotfmr} and magnetic switching.\cite{staszek2019,ohnoswitching,garbandela2014,lau2016spin} Especially for the F layers with a magnetic perpedicular anisotropy, the SOT enables a promising way to design efficient, ultra-low power consumption spintronic devices. Apart from the SHE and related spin accumulation, the other pure interfacial effects, like charge-spin conversion REE at interfaces, contribute to the SOT.\cite{REESHE,lorenz2014,kang2019} Therefore the interface engineering becomes significant for optimization of SOT-based devices.\cite{inzynier2019,beach2019,buhrman2019} The spin-currents injected to F layer and SOTs may be examined by electric measurements through its magnetoresistance\cite{karwacki2020,skowronski2016temperature,cecot2017influence} and the Anomalous Hall effect (AHE).\cite{yamagouchiahe} The change of the resistance of the hybride structure caused by the above effects are refered to as spin magnetoresistace (SMR).\cite{kim} Up to date, the specific multilayer structures (bi- and trilayers) were studied in details. Among them we find CoFeB-based structures like W/i-CoFeB/Pt,\cite{witek2019} p-CoFeB/Ta, as well as the Co-based multilayers like \cite{ralph2012} p-Co/Pt/i-Co,\cite{staszek2019} p-Co/Pt,\cite{avcicrystal,hayashi2021spin,avci2015unidirectional} i-Co/Ta,\cite{avci2015unidirectional} Ta/i-Co/Pt,\cite{luo2019dependence} where p(i) stands for perpendicular(in-plane) anisotropy. In this paper we present the detailed studies of the Co/Pt/Co system with use of the electrically-detected FMR (ferromagnetic resonance) as well as low-frequency harmonic Hall voltage and static magnetotransport measurements. 
We provide the results on the resonance frequencies and the SOT effective fields depending on the Pt thickness. Also, we analyse the magnetic parameters of the system like anisotropies, saturation magnetizations and the interlayer exchange coupling (IEC). We show that anisotropies and the IEC strongly depend on the Pt thickness, particularly for Pt layer thicknesses less than 2 nm. For such a thin Pt, the transition from the effective in-plane Co anisotropy to the perpendicular one may occur. We account for the features by providing a reliable theoretical macrospin models of magnetization dynamics, magnetoresistance and effective spin Hall effect angle.

\section{Experimental\label{sec:exp}}
\subsection{Sample preparation}

\begin{figure}[H]
\includegraphics[width=8.5cm]{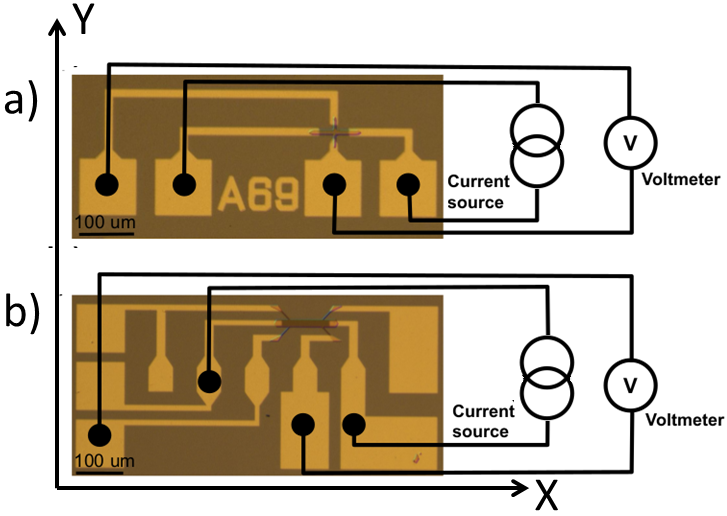}
\caption{The Hall resistance ($R_{xy}$) (a) and 4-points longitudinal magnetoresistance (b) measurement setups depicted on micrographs of the devices.}
\label{fig:schemeMR}
\end{figure}

Multilayers are deposited on thermally oxidized Si substrates using magnetron sputtering at room
temperature. We study Co/Pt/Co trilayer within Ti(2)/Co(1)/Pt(0-4)/Co(1)/MgO(2) /Ti(2) structure (the numbers in brackets indicate the nominal thickness of the individual layers in nanometers).
\begin{figure}[ht]
\includegraphics[width=8.5cm]{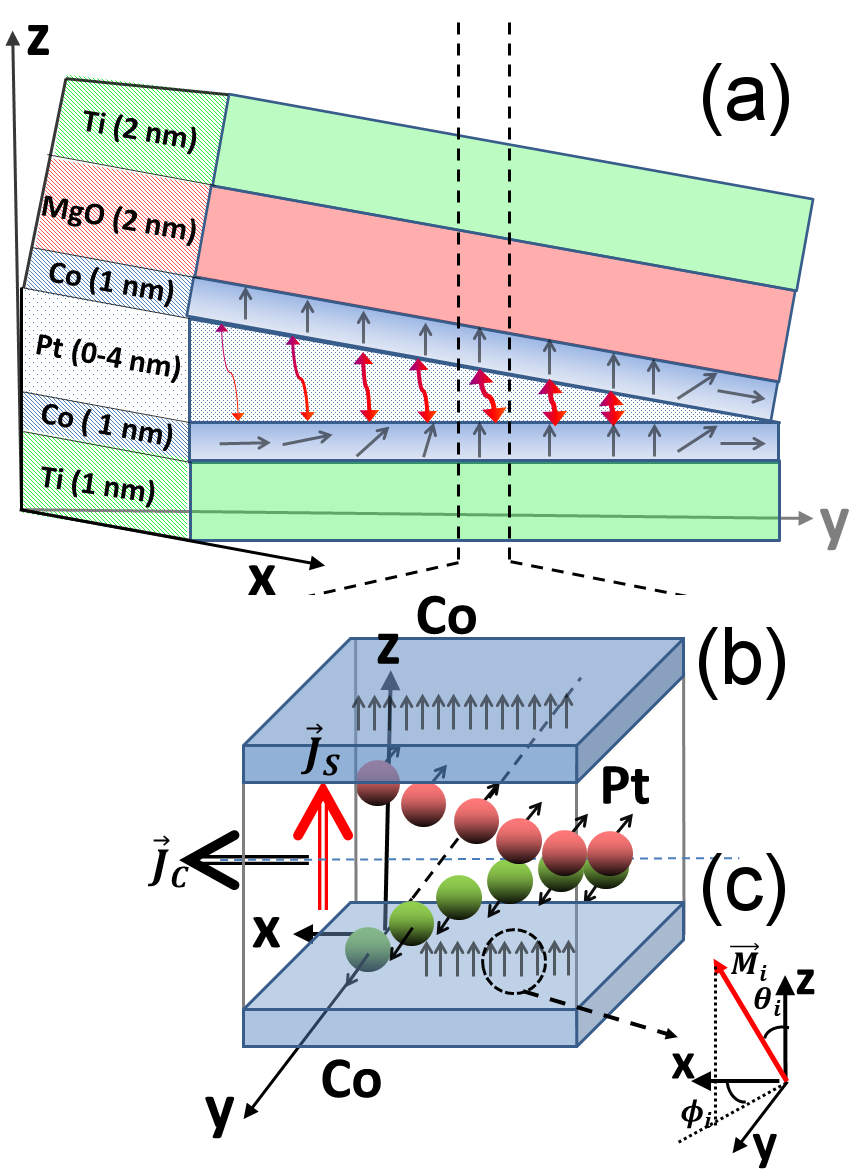}
\caption{(a) Experimental multilayer stack with a wedge of Pt. The red thick (thin) wavy arrows indicate strong (weak) IEC, whereas the grey arrows show the change in the magnetization alignment with Pt thickness, (b) the patterned device for a certain thickness of Pt - the arrows indicate the direction of current flow ($\vec{j_c}$) and associated spin current ($\vec{j_s}$) due to SHE. The short arrows depicted in Co layers mean their magnetization vectors for a given Pt thickness at remanence, (c) the polar and azimuthal angles describing the magnetization direction within the Co layers.}
\label{fig:schemat}
\end{figure}
The Co/Pt/Co trilayer was designed so it allows to study the influence of Pt thickness on the magnetic anisotropy of bottom and top Co layers, the IEC between Co layers through Pt spacer,  magnetization dynamics, and SHE-driven SOT acting on the F layers. For this purpose, both bottom and top thin Co layers should have small anisotropy (differing by interfaces Ti/Co and Co/MgO) with values close to the transition from in-plane to perpendicular. The Ti underlayer improves subsequent layers' adhesion and smooths the substrate surface. Moreover, as shown in Ref.\citenum{sule2008asymmetric}, Ti/Co interface is alloyed due to mixing during magnetron deposition, while Co/MgO interface is sharp.\cite{gweon2018} Therefore, the top Co layer is characterized by higher interface perpendicular anisotropy. The Pt layer was deposited in wedge-shaped form with thickness varied from 0 to 4 nm along 20 mm long sample edge (x coordinate). The resulting thickness gradient was achieved by the controlled movement of a shutter. Thicknesses of all layers were determined from the deposition growth rate of particular materials calibrated using x-ray reflectivity measurements. Next, before patterning to the form of bar devices, all as-deposited samples were characterized by x-ray diffraction $\theta-2\theta$ (XRD) and grazing incidence diffraction (GIXD),  and also examined by the polar Kerr magnetometer (p-MOKE) and time-resolved TR-MOKE in order to determine the static and dynamic magnetization parameters, which studies have been described in detail in a separate work.\cite{bonda2020} After basic characterization of continuous samples, multilayers were patterned using optical direct imaging lithography and ion etching to create a matrix of Hall- and resistance-bar devices, with different thicknesses of Pt for subsequent electrical measurements (Figure \ref{fig:schemat}a). The sizes of prepared structures were 100$\mu$m x 20$\mu$m for magnetoresistance and spin diode effect measurements, whereas 100 $\mu m$ x 10 $\mu m$ for the AHE, and harmonics measurements. Al(20)/Au(30) electrical leads of 100 $\mu$m x 100 $\mu$m were deposited in a second lithography step followed by the lift-off process. Specific locations of pads near the Hall-bars were designed for measurement in a custom-made rotating probe station allowing 2- or 4-points measurement of electrical transport properties in the presence of the magnetic field applied at an arbitrary azimuthal and polar angle with respect to the Hall-bar axis. The scheme of experimental setup for longitudinal ($R_{xx}$) and Hall ($R_{xy}$) resistance measurements is shown in Figure \ref{fig:schemeMR}.
The resistance was measured using a four-point method\cite{4points} and resistivities of Pt and Co layers were determined using a parallel resistors model and the method described by Kawaguchi et al.\cite{kawaguczi}. The Pt and Co resistivities analysis yielded 59$\mu\Omega$ cm and 72.5 $\mu\Omega$ cm, respectively.

\subsection{Structural characterization}
\begin{figure}[H]
\includegraphics[width=8.5cm]{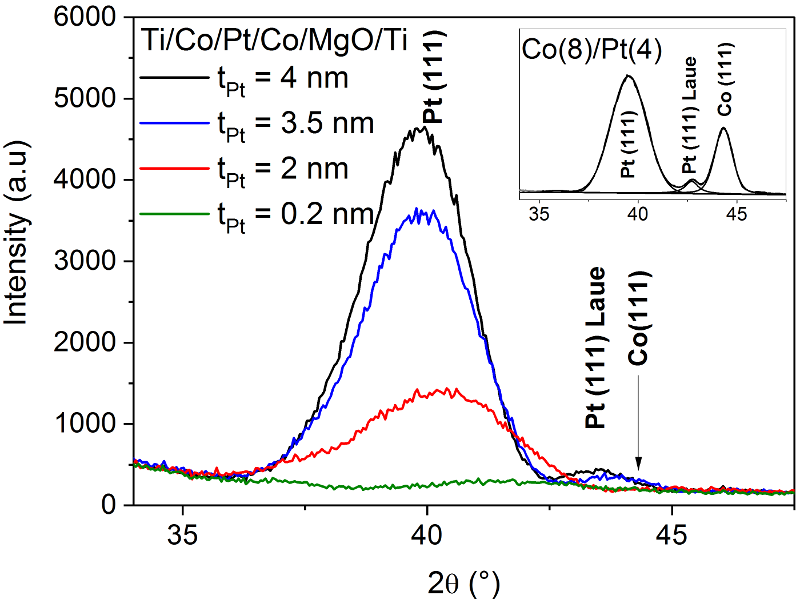}
\caption{The XRD $\theta-2\theta$ profiles of a Si/SiO2/Ti/Co(1)/Pt(0-4)/Co/MgO/Ti measured at different position of the Pt wedge. The arrow indicates $2\theta$ position of the structural Co (111) peak visible in the reference sample with 8 nm of Co and 4 nm of Pt layer thickness (inset).}
\label{fig:jk2}
\end{figure}

High-resolution X’Pert–MPD diffractometer with a Cu anode was used for x-ray diffraction (XRD) characterization.
Figure \ref{fig:jk2} shows the XRD $\theta-2\theta$ profiles of the Si/SiO2/Ti/Co(1)/Pt(0-4)/Co(1)/MgO/Ti multilayer measured at different positions of the Pt wedge. The $\theta-2\theta$ measurements show the preferred growth of the Pt/Co in [111] direction of fcc structure. The peak of Co layers is invisible because of their tiny thicknesses ($t_{Co} \approx 1$ nm).
The arrows indicate the Co (111) peak position present in the thick Co layer case (see supplemental material in Refs.\citenum{staszek2019,kanak2013x}). Peak on the right side of the Pt(111) is a Laue satellite\cite{staszek2019} that confirms the asymmetry in top Pt/Co and bottom Co/Pt interfaces. The intensity of the profiles depends on the number of Pt atoms. Therefore, in the case of very thin (0.2 nm) Pt layers, the Laue peaks are out of detection of the experimental method. However, one can see that the Pt peak slightly shifts to the right for thin Pt layers. It suggests that most of the Pt layer becomes mixed with Co atoms making a rather Co-Pt compound than a well-separated layer.

\begin{figure}[H]
\includegraphics[width=8.5cm]{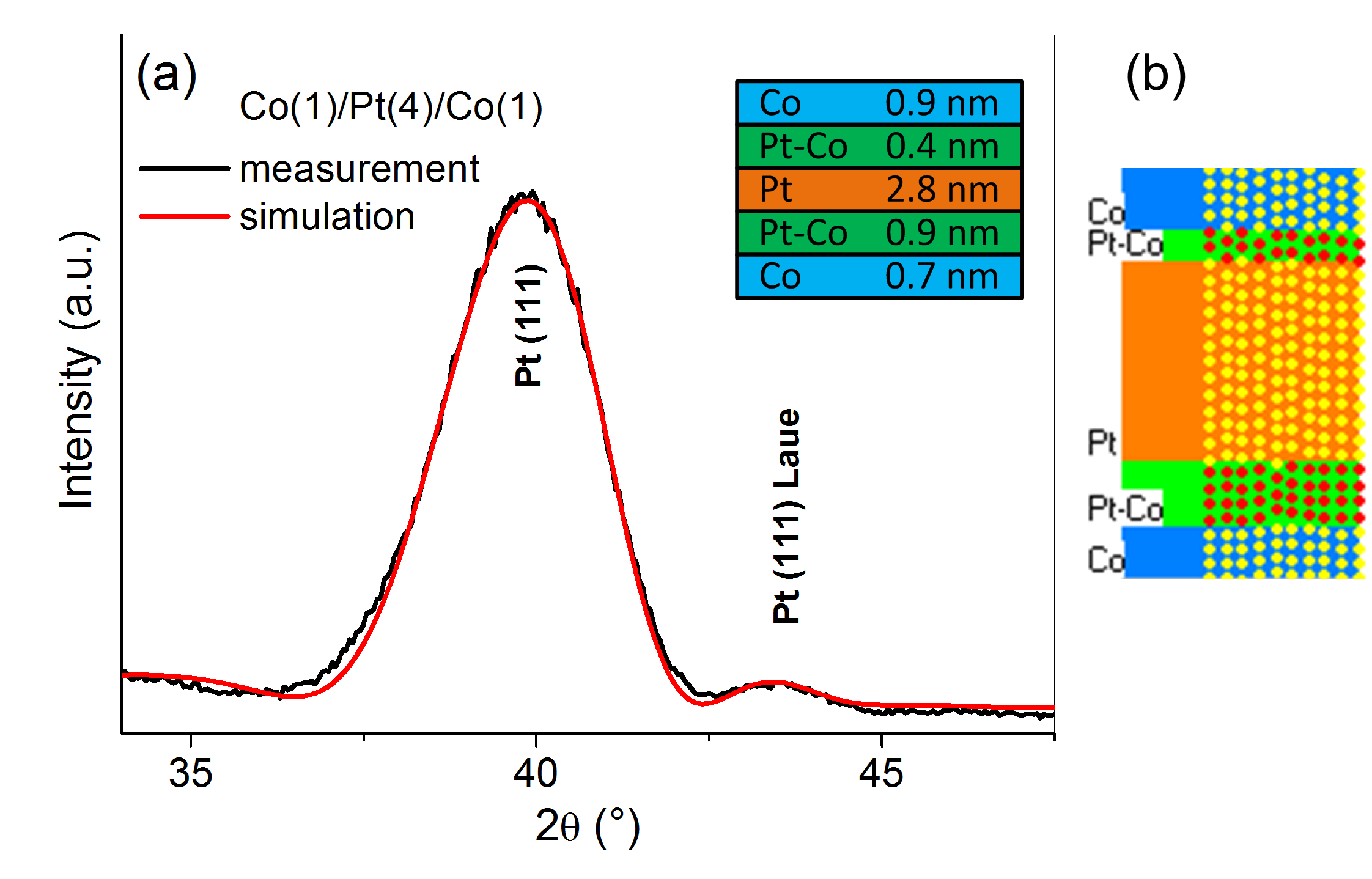}
\caption{Measured and calculated XRD $\theta-2\theta$ profiles (a). The assumed thicknesses of the Pt, Co layers and transition area with the Pt-Co mixed interfaces (inset). A snapshot of the Monte Carlo simulation of the interface structure in Co/Pt/Co (b).}
\label{fig:jk3}
\end{figure}

Figure \ref{fig:jk3}a shows the profile measured for the Si/SiO2/Ti(2)/Co(1)/Pt(4)/Co(1)/MgO(2) /Ti(2) at thick Pt $\approx$ 4 nm, together with the profile calculated using the simulated\cite{kanak2013x} structure (Figure  \ref{fig:jk3}b). An excellent agreement between the experimental and theoretical values is achieved. The structure was simulated with the assumption of Pt and Co mixing at the interfaces. The simulations represent the columnar grains in the Pt and Co layers and a transition area with the Pt-Co mixed interface. Mixing of the Pt and Co atoms at interface causes decreasing Pt lattice planes spacing compared to pure Pt.

The simulation assumes more significant mixing at the bottom side of the Pt than at the top one. In the former case, the heavy Pt atoms can penetrate the Co layer more easily than in the latter case. Moreover, for Pt within Co, the interfacial enthalpy is -33 kJ/(mole of atoms), whereas for Co in Pt is -26 kJ/(mole of atoms). Higher negative enthalpy results in easier mixing at the bottom Co/Pt  interface.

\subsection{Spin diode effect measurements\label{sec:sde}}
\begin{figure}[H]
\includegraphics[width=8.5cm]{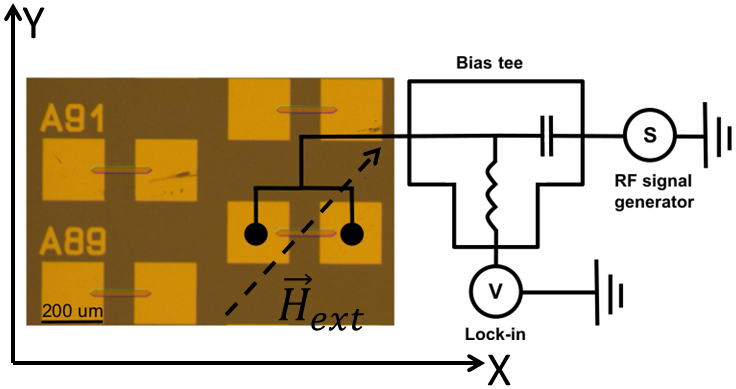}
\caption{The resistance-bar attached to the spin-diode FMR measurement setup (see main text for details) depicted on the set of four patterned devices photomicrography. The dashed arrow indicates the direction of the external magnetic field $\vec{H}_{ext}$.}
\label{fig:sdescheme}
\end{figure}

The magnetic dynamics of the patterned samples was electrically-detected with the FMR measurements through the spin diode effect.\cite{tulapurkar2005} The scheme of measurement setup is shown in Figure \ref{fig:sdescheme}. The effect occurs when the rf current flows through the
magnetoresistive element that in the case of our samples exhibits the anisotropic magnetoresistance (AMR) and SMR. Then, the current-related effective magnetic fields (as Oersted field or SOT fields) force the sample magnetization to oscillate. The magnetization oscillations, in turn, result in time-dependent resistance of the sample which mixes up with the rf-current.

Therfore the measured output voltage may be expressed as $V_{out} = I_0 \cos (\omega t ) \cdot  R(\omega t + \Phi)$, where the $\Phi$ is the phase shift between the current and resistance.  
One notes that $V_{out}$ includes ac and dc  contributions, namely: $V_{out} = V_{dc} + V_{ac}= I_0 \delta R \cos\Phi +  I_0  \delta R(2 \omega t + \Phi)$.  The dc output voltage depends on the angular frequency, external magnetic field and parameters of the sample. 

In this paper, the spin-diode FMR measurements are performed with an amplitude-modulated radio-frequency (rf) current with a corresponding power of P = 16 dBm and frequencies
ranging from 1 to 25 GHz.  The mixing voltage ($V_{out}$) is measured using a lock-in amplifier synchronized to the rf signal. The in-plane magnetic field ($H_{ext}$) is applied at $\phi$ = 45 deg with respect to the microstrip long axis and was swept from 0 up to 9 kOe. 

\subsection{Harmonic measurements\label{sec:harmonic}}
\begin{figure}[H]
\includegraphics[width=8.5cm]{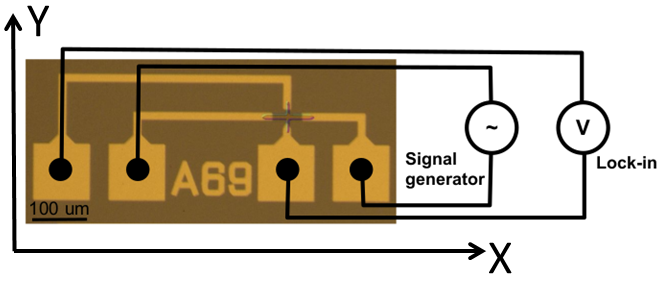}
\caption{The Hall-bar and the experimental setup for harmonic Hall voltage ($V_{\omega,2\omega}$) measurements.}
\label{fig:harmonicsscheme}
\end{figure}

In order to determine spin-orbit torque fields (damping- and field-like components), as well as the spin Hall angle, we used the methods based on the harmonic measurements.\cite{avci2014,Lau_2017,meinert}
For these measurements, we apply a low-frequency constant-amplitude sinusoidal voltage to the Hall bar device with current density from j= $5.75 \times 10^{9}$ $A/m^2$ to j= $9.62 \times 10^{9}$ $A/m^2$ depending on the Pt layer thickness. Using two lock-in amplifiers, we measure simultaneously the in-phase first harmonic ($V_\omega$) and the out-of-phase second harmonic Hall voltages ($V_{2\omega}$) as a function of an external magnetic field $H_{ext}$. The sample is rotated within the x-y plane, making an azimuthal angle $\phi_H$ with the x-axis as depicted in Figure \ref{fig:harmonicsscheme} on which the scheme of measurement setup is shown.
The measurements were conducted in two configurations: the first one is referred to as field measurements, and the samples were probed with the different magnitudes of the external magnetic field applied along both x and y directions.\cite{hayashi2014}, while the second configuration is the angular measurements. The sample is rotated in x-y plane while the $V_{\omega,2\omega}$ is recorded \cite{hayashi2014,avci2014} for fixed magnitudes of external magnetic field. The field measurements are relevant in the case of samples with out-of-plane effective anisotropies. On the contrary, the angular measurements allow to detect harmonic signals in samples with in-plane effective anisotropy. 

\section{Theory\label{sec:th}}
\subsection{Resonance model\label{sec:eigenmodes}}
This subsection presents the macrospin model that allow us to calculate resonance frequencies of the considered Co/Pt/Co structure. 
Since the Co layers may be either coupled or uncoupled we employ the approach that has been already presented in details and successfully applied in our previous work.\cite{ogrodnikPRM} We describe magnetic moments of each layer by spherical angles (polar \(\theta_i\) and azimuthal \(\phi_i\)): 
\begin{equation}
\bm{M_{i}} = M_{S,i} [ \sin\theta_{i} \cos\phi_{i},\, \sin \theta_{i} \sin\phi_{i} ,\, \cos\theta_{i}],
\end{equation}
where $i$=1(2) is reffered as to bottom(top) cobalt layer. 
The magnetization dynamics of the system is described by two coupled Landau--Lifshitz--Gilbert (LLG) equations,
\begin{eqnarray}
\label{eq:LLGset}
\frac{d\bm{M_i}}{dt} = -\gamma_e \bm{M_i} \times \bm{H}_{\rm eff,i} + \frac{\alpha_g}{M_{S,i}} \bm{M_i} \times \frac{d\bm{M_i}}{dt} + \nonumber \\ 
+ \gamma_e (\bm{\tau}_{DL,i} + \bm{\tau}_{FL,i}), 
\end{eqnarray}
where \(\gamma_e\approx1.760859644\times 10^{11}\ \frac{\mathrm{rad}}{\mathrm{sT}}\) is the gyromagnetic ratio, and \(\alpha_g\)  is the Gilbert damping parameter for each layer.

The terms $\bm{\tau}_{DL} = H_{DL} ( \bm{m}_i \times \bm{m}_i \times \bm{\hat{e}}_y )$ 
and $\bm{\tau}_{FL} = H_{FL} ( \bm{m}_i \times \bm{\hat{e}}_y )$ stand for SOT damping-like (DL) and field-like(FL) components with the unit vector $\bm{m_{1(2)}} = \frac{\bm{M_{1(2)}}}{M_{S,1(2)}}$ and the amplitudes $H_{DL}$ and $H_{FL}$ respectively. 


The effective field ($H_{eff}$) can be expressed as functional derivative of the follwing total magnetic energy of the system:
\begin{eqnarray}
U =  \sum_{i=1}^2 
K_{\perp,i} t_{Co_i} \left[ \left( \cos\beta_i \sin \theta_i \sin \phi_i - \sin \beta_i \cos\theta_i \right)^2+ \right.\nonumber \\
+ \left( \cos\delta_i \left( \cos\theta_i \cos \beta_i  + \sin \theta_i \sin \phi_i \sin \beta_i \right) - \right. \nonumber \\
\left.\left. -\cos \phi_i \sin \theta_i \sin \delta \right)^2 \right]+ K_{\parallel ,i} \left( \cos^2\theta_i + \sin^2 \theta_i \sin^2 \phi_i \right)- \nonumber\\
-t_{Co_i} \bm{M_i} \cdot \bm{H}_{\rm ext} -t_{Co_i} \bm{M_i} \cdot \bm{H}_{dem,i} -  J \bm{M_i} \cdot \bm{M_j}, 
\label{eq:energia}
 \end{eqnarray}
The complex expression for the anisotropies originates from the rotation of the easy-axes around x and y directions with use of the relevant Euler rotation matrices.
The angles $\beta$ and $\delta$ have been introduced in order to account for a small deviation of the perpendicular anisotropies($K_{\perp,i}$) from perpendicular(z) direction ($\delta,\beta\ll \pi/2$). 
The perpendicular anisotropy terms simplify into well-known form $K_{\perp,i} \sin^2\theta_i$  when $\delta_i,\beta_i = 0$. 
Also, we have added a small in-plane contribution $K_\parallel \ll K_\perp$ along $y$ direction. As long as they are small, they slightly improve the fitting the macrospin model to the experimental data. 
In Eq.(\ref{eq:energia}), $t_{Co,i}$, $\bm{H}_{\rm ext}$, $\bm{H}_{dem,i}$ and  $J$ stand for magnetic layer thickness, external magnetic field, demagnetizing field and IEC respectively.  

The LLG equation (\ref{eq:LLGset}) in polar coordinates can be written in the general form
\begin{equation}
\dot{\bm{\alpha}} = \bm{v}(\theta_i,\phi_i)^T,
\label{eq:llgpolar}
\end{equation}
where $\dot{\bm{\alpha}}$ and \(\bm{v}\) are the vectors containing the spherical angles ($\theta_{1,2},\phi_{1,2}$) time-derivatives and  the right-hand-side (RHS) of the LLG equation, respectively. After linearization of \(\bm{v}\) with respect to small deviations in \(\theta_i\) and \(\phi_i\) from their stationary values, one can write  Eq. (\ref{eq:llgpolar}) in the form
\begin{equation}
\dot{\bm{\alpha}} = \hat{X}  \bm{\Gamma}(t) ,
\label{eq:llglinear}
  \end{equation}
where  \(\hat{X}\) is a \(4 \times 4\) matrix consisting of the derivatives of the RHS of Eq. (\ref{eq:llgpolar}) with respect to the angles \(\theta_i,\,\phi_i\) (i.e., \(X_{kj} \equiv \frac{\partial v_{k}}{\partial \alpha_j}\)), while \(\bm{\Gamma}(t)= \bm(\delta\alpha_1(t), \ldots, \delta \alpha_{4}(t) \bm )^T\) is a vector containing time-dependent angle differentials, i.e. \(\delta \alpha_1(t) \equiv \delta \theta_1(t)\), \(\delta \alpha_2(t) \equiv \delta \phi_1(t)\), etc.
When small oscillations assumed and in the absence of the external driving force (i.e. SOT or Oersted field)  Eq. (\ref{eq:llglinear}) can be rewritten as an eigenvalue problem of the matrix \(\hat{X}\):
 \begin{equation}
  \left| \hat{X} - \omega \hat{I} \right| = 0.
  \label{eq:czesto}
 \end{equation}
The solution of the problem provides the complex eigenvalues \(\omega_i\) determine 2 distinct natural resonance angular frequencies of the system, \(\omega_{R,i} = \text{Re}\, \omega_i \). 

\subsection{Diffusive model of the magnetoresistance\label{sec:lukasz}}

The average longitudinal resistance of our trilayer stack is in general dependent on the orientation of magnetizations $\bm{m}_{1(2)}$ 
in both ferromagnetic layers and reads
\begin{align}
    \label{eq:MR}
R_{xx}(\bm{m_{1(2)}})=\frac{L}{w}\left[\frac{1}{E_x}\frac{1}{\sum_\chi t_{\chi}}\sum_{\chi}\int_{\chi} dz j_{c,x}^{\chi}(z,\bm{m_{1(2)}})\right]^{-1}\,,
\end{align}
where $E_x$ is the electric field in $x$ direction, $L$ is the length, $w$ the width, and $t_{\chi}$ the thickness of layer $\chi=HM, F1, F2$ and $\int_{\chi}$ denotes integral with limits corresponding to the position of layer $\chi$ in the stack. For Pt, i.e. for $\chi=HM$, the charge current density reads
\begin{align}
j_{c,x}^{HM}(z,\bm{m_{1(2)}})=\frac{1}{\rho_{HM}}E_x-\frac{\theta_{SH}}{2e\rho_{HM}}\frac{\partial \mu_{s,y}^{HM}(z,\bm{m_{1(2)}})}{\partial z} \,,
\label{eq:jchm}
\end{align}
where $\rho_{HM}$ is the resistivity of the Pt layer, $\theta_{SH}$ is the spin Hall angle defined as charge-to-spin current conversion efficiency at a very thick HM layer limit, and $\mu_{s,y}^{HM}(z,\bm{m_{1(2)}})$ is the spin accumulation,
while for the ferromagnetic layers, i.e. $\chi=F1$ ($\chi=F2$)
\begin{align}
j_{c,x}^{F1(F2)}(z,\bm{m_{1(2)}})=\frac{\left[1-\theta_{AMR}(\bm{m_{1(2)}}\cdot\hat{\bm{x}})^2\right]}{\rho_{F1(F2)}}E_x \,,
\label{eq:jcf}
\end{align}
where $\rho_{F1(F2)}$ is the resistivity of the corresponding ferromagnetic layer, $\theta_{AMR}$ is the AMR in the thick ferromagnet limit (assumed for simplicity the same in both ferromagnetic layers). For more details see e.g.~\citenum{karwacki2020}.

To obtain spin accumulation in the Pt layer we consider the spin current density flowing in Pt,
\begin{align}
\mathbf{j}_{s}^{HM}(z,\bm{m_{1(2)}})=-\theta_{SH}\frac{1}{\rho_{HM}}E_x\hat{\bm{y}}+\frac{1}{2e\rho_{HM}}\frac{\partial \bm{\mu}_s^{HM}(z,\bm{m_{1(2)}})}{\partial z} \,,
\end{align}
along with the boundary conditions:
\begin{subequations}
\begin{align}
\mathbf{j}_s^{HM}(-t_H,\bm{m_1})&=\mathbf{q}_{1}(\bm{m_1})\,, \\
\mathbf{j}_s^{HM}(0,\bm{m_2})&=-\mathbf{q}_2(\bm{m_2})\,, \\
j_s^{F1}(-t_{HM}-t_{F1})&=0\,,  \\
j_s^{F1}(-t_{HM})&=\mathbf{m_1}\cdot \mathbf{q}_1(\bm{m_1})\,, \\
j_s^{F2}(0)&=\mathbf{m_2}\cdot\mathbf{q}_2(\bm{m_2})\,,  \\
j_s^{F2}(t_{F2})&=0\,,
\end{align}
\end{subequations}
where the spin current in ferromagnetic metals assumes the following form
\begin{equation}
j_{s}^{F1(F2)}(z)=\frac{(1-\beta_{F1(F2)}^2)}{2e\rho_{F1(F2)}}\frac{\partial \mu_s^{F1(F2)}(z)}{\partial z}\,,
\end{equation}
and the interfacial spin currents:
\begin{subequations}
\begin{align}
\bm{q}_{1}(\bm{m_1})&=G_s^{(1)}\left[\mu_s^{F1}(-t_{HM})-\bm{m_1}\cdot\bm{\mu}_s^{HM}(-t_{HM},\bm{m_{1(2)}}) \right]\bm{m_1}\nonumber \\
&+G_r^{(1)}\bm{m_1}\times\mathbf{m_1}\times\bm{\mu}_s^{HM}(-t_{HM},\bm{m_{1(2)}}) \nonumber \\
&+G_i^{(1)}\bm{m_1}\times\bm{\mu}_s^{HM}(-t_{HM},\bm{m_{1(2)}}) \\
\bm{q}_{2}(\bm{m_2})&=G_s^{(2)}\left[\mu_s^{F2}(0)-\bm{m_2}\cdot\bm{\mu}_s^{HM}(0,\bm{m_{1(2)}}) \right]\bm{m_2}\nonumber \\
&+G_r^{(2)}\bm{m_2}\times\bm{m_2}\times\bm{\mu}_s^{HM}(0,\bm{m_{1(2)}}) \nonumber \\
&+G_i^{(2)}\bm{m_2}\times\bm{\mu}_s^{HM}(0,\bm{m_{1(2)}}) \,,
\end{align}
\end{subequations}
where $G_s^{(1)}$ and $G_{s}^{(2)}$ are spin conductances and $G_{r(i)}^{(1)}$ and $G_{r(i)}^{(2)}$ are real (imaginary) parts of spin-mixing conductances for interface 1 (F1/HM) and 2 (HM/F2), respectively.
Moreover, the effective fields: $H_{DL}$ and $H_{FL}$ (cf. Sec.\ref{sec:eigenmodes}) due to SHE and spin accumulation at the interfaces can be expressed as follows:
\begin{equation}
H_{DL}^{1(2)} = -\frac{\hbar}{2e^2}\frac{1}{\mu_0 M_{S,1(2)} t_{F1(F2)} } \mathbf{x} \cdot (\mathbf{m_{1(2)}} \times \mathbf{q_{1(2)}} )
\label{eq:HDL}
\end{equation}
and
\begin{equation}
H_{FL}^{1(2)} = -\frac{\hbar}{2e^2}\frac{1}{\mu_0 M_{S,1(2)} t_{F1(F2)} } \mathbf{y} \cdot (\mathbf{m_{1(2)}} \times \mathbf{q_{1(2)}} ).
\label{eq:HFL}
\end{equation}
To fit the appropriate magnetoresistance relations obtained with use of Eq.~(\ref{eq:MR}) and shown in Figure \ref{fig:figura3}, we use the following parameters~\cite{karwacki2020}: $\rho_{HM}=59~\mu\Omega\text{cm}$, $\rho_{F1(F2)}=72.5~\mu\Omega\text{cm}$, $\lambda_{HM}=1.8~\text{nm}$, $\lambda_{F1}=\lambda_{F2}=7~\text{nm}$, $\theta_{SH}=8\%$, $\theta_{AMR}=0.15\%$, $\beta_1=\beta_2=0.3$, $G_s^{(1)}=G_s^{(2)}=G_r^{(1)}=G_i^{(1)}=10^{15}~\Omega^{-1}m^{-2}$, $G_r^{(2)}=G_i^{(2)}= 0.4 G_r^{(1)}$ The parameters were also used to calculate SOT effective fields that turn out to be pivotal in the interpretation of the experimental data presented in Sec.\ref{sec:SOT}. 

\section{Results\label{sec:res}}
\subsection{FMR and interlayer coupling  \label{sec:FMR}}
\begin{figure}[H]
\includegraphics[width=8.5cm]{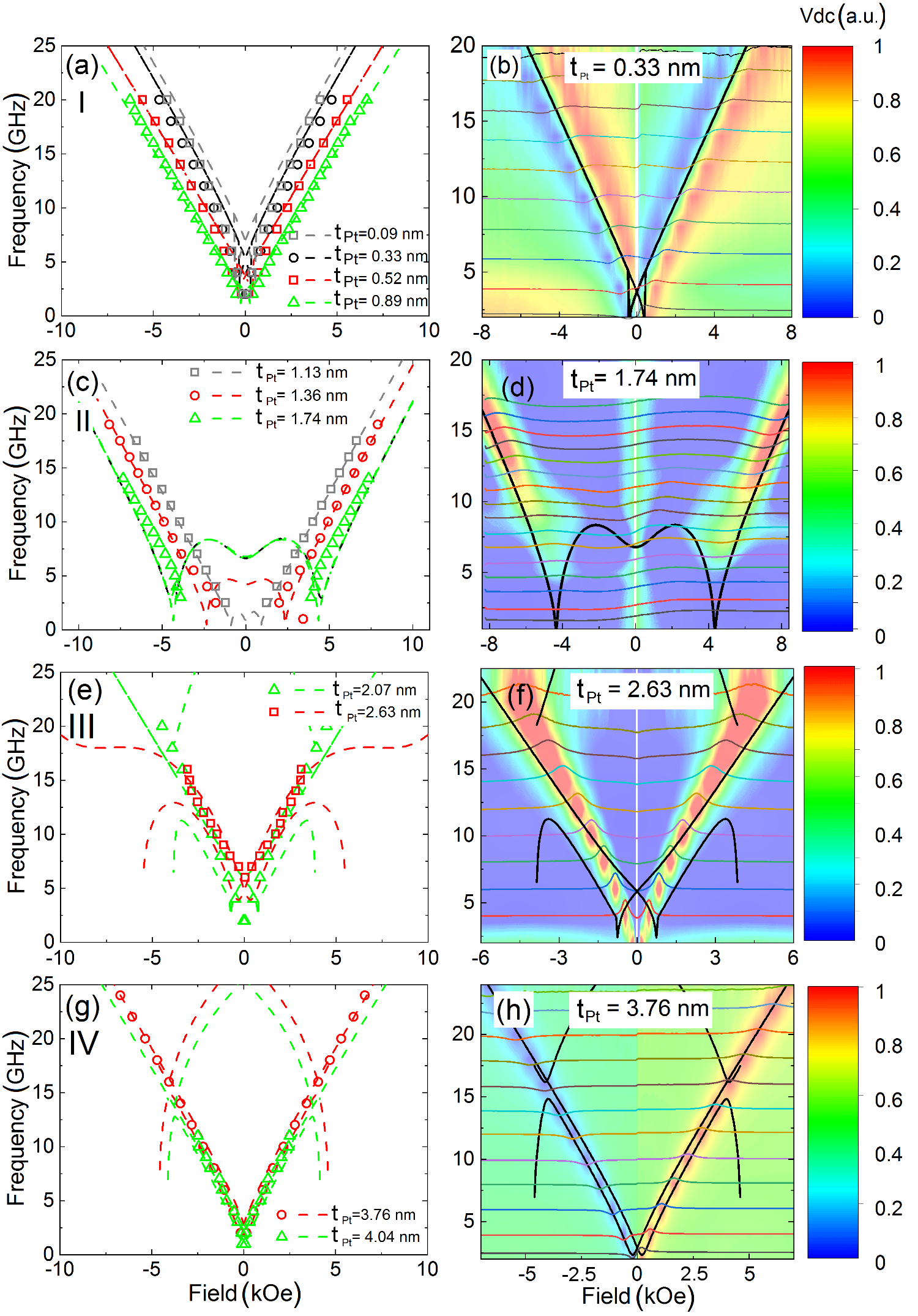}
\caption{Experimental versus theoretical relations of dispersion for samples from region I ((a) and (b)), II (c-d), III (e-f), IV (g-h). Left column: the sets of  theoretical (lines) and experimental (points) dependencies for each region. Right column: the experimental $V_{DC}$ spectra shown as color map (color is the magnitude of SD signal) and the source raw spectra measured at frequency ranging from 2 to 20 GHz (light color lines) with corresponding theoretical f(H) dependencies (solid black lines). The macrospin simulation magnetic parameters are the same as presented in Figure \ref{fig:parametry}(a-d).  }

\label{fig:figura0}
\end{figure}

First, we measured the magnetization dynamics of the Co/Pt/Co sample in a wide range of Pt thickness from strong through moderate coupling to completely decoupled Co layers. The dynamics was investigated using the electrically detected FMR through the spin diode effect\cite{tulapurkar2005} as described in Sec.\ref{sec:sde}. We observed the dispersion relations changing once the Pt thickness reaches boundary values. Thus, we could point out the distinct regions where the system behaves differently. This feature
is illustrated in Figure \ref{fig:figura0}. On the right panel of Figure \ref{fig:figura0} one can see the color FMR spectral line-shapes. On the left panel, points correspond to the experimental resonance frequencies. On both sides of this figure, we show that for thin Pt spacer (below 1 nm), the dispersion relations are typical Kittel-like dependencies and move towards lower frequencies when the $t_{Pt}$ is growing. Next, for $t_{Pt} > 1 nm$, the $f(H_{r})$ change their slopes. Also, their branches part from each other, especially at low frequencies when a sort of resonance mode gaps in experimental data occur. The increase of the Pt thickness ($t_{Pt}$ > 2 nm) provides the Kittel-like dependencies again. However, for thick Pt ($t_{Pt}$>3 nm) the experimental $f(H_r)$ practically does not change anymore with the Pt thickness.

In order to understand the dependence on Pt thickness, we did the macrospin simulations. We performed a vast number of fittings for the whole range of the Pt thickness, i.e. from 0.09 nm to 4.04 nm (Figure \ref{fig:figura0}). We reproduced the difference in the dynamical behavior of the structure, and we were able to identify the boundaries between different regions of Pt thickness where these behaviors occur, namely region I (thin Pt), II (medium Pt), III (intermediate Pt) and region IV (thick Pt). Despite the similarity of $f(H_r)$ in regions III and IV, we refer to the former as the intermediate since the differences occur in magnetoresistance results discussed in the further part of the paper.
The presence of the additional modes (especially in regions III and IV) that were not registered in the experiment, we explain as follows. The theoretical results based on the model presented in Sec.\ref{sec:eigenmodes} come from the solution of the eigenvalue problem, i.e., the model predicts all possible steady-state modes, regardless of the source of their excitations. On the other hand, the different way of forcing the excitation (by SOTs or Oersted field) is inherent. In general, the resonance modes are not always exicted, depending on the force amplitude and its origin.

\begin{figure}[H]
\includegraphics[width=8.5cm]{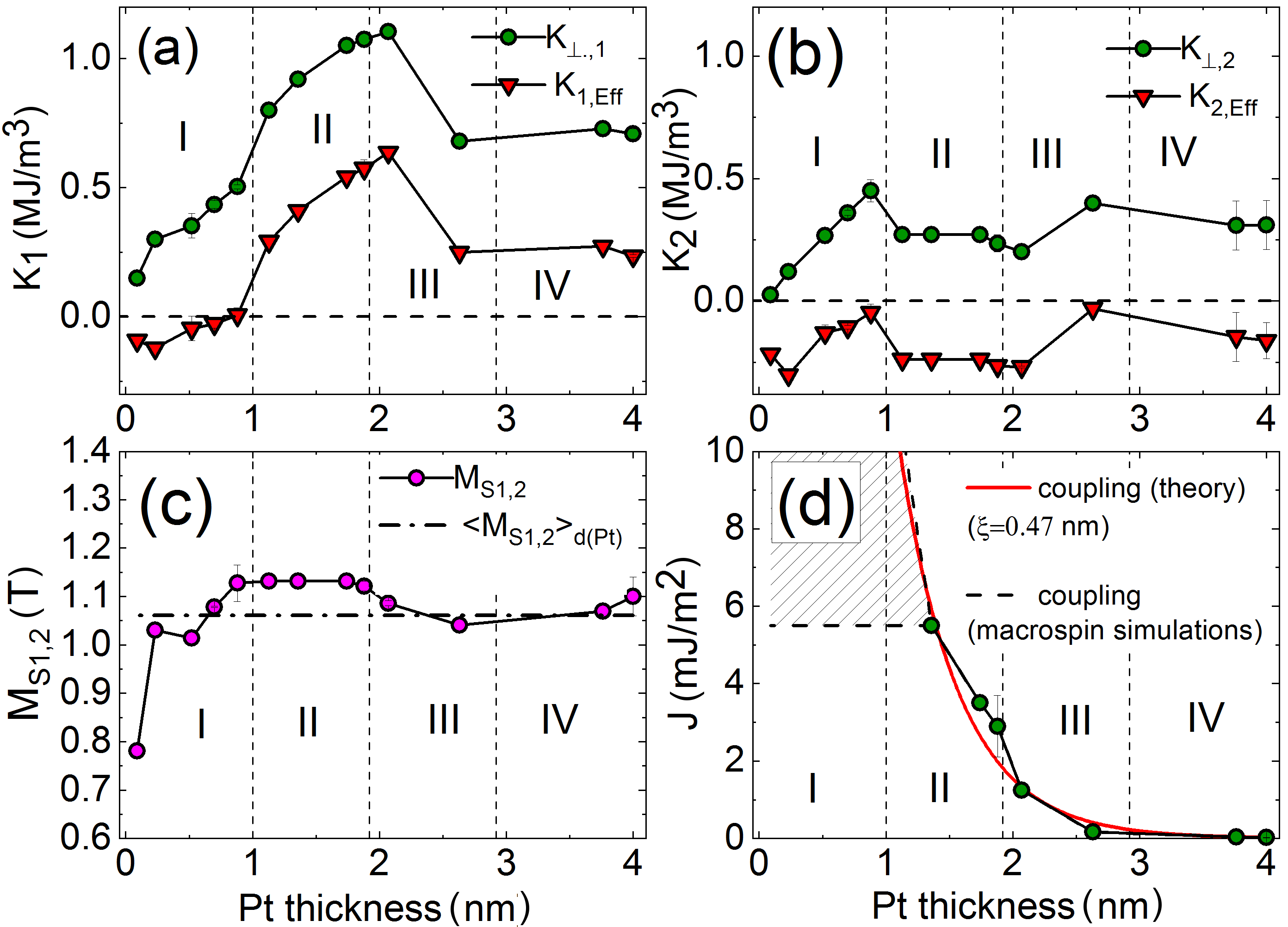}
\caption{Magnetic parameters of Co layers as a function of Pt layer thickness derived from the macrospin simulations of the spin-diode FMR spectra: perpendicular and effective anisotropies (a) and (b), magnetization saturation (c) and interlayer coupling (d).}
\label{fig:parametry}
\end{figure}

The macrospin parameters are summarized in Figure \ref{fig:parametry}. The vertical dashed lines indicate the boundaries between regions I, II, III, and IV. Figure \ref{fig:parametry}(a-d) presents: perpendicular anisotropies $K_{\perp,1,2}$, saturation magnetization $M_{S,1,2}$, as well as the strength of the interlayer coupling. We also depicted the effective anisotropies $K_{eff,1,2} \equiv K_{\perp} - \frac{\mu_0}{2} M^2_S $. One can see that the Co layer (indexed as 1) covered by the MgO layer exhibits a larger perpendicular anisotropy than the one adjacent to the Ti layer (indexed as 2), similarly as in the system Si/SiO2/Ti(2)/Co(3)/Pt(t$_{Pt}$)/Co(1)/MgO(2)/Ti(2) examined in our previous work.\cite{staszek2019} Moreover, on the basis of magnetization measurements in an external perpendicular field, using vibrating sample magnetometer (VSM), we showed that sample Pt(4)/Co(1)/MgO(2)/Ti(2) has smaller effective anisotropy field ($H_{K,eff}$=1.3 kOe) than Ti(2)/Co(1)/Pt(4) ($H_{K,eff}$=1.65 kOe). In Figure \ref{fig:parametry}a,b we show that the effective anisotropy $K_{eff,1}$ changes its sign, while the $K_{eff,2}$ is negative for all Pt thicknesses. The change of the sign of the effective anisotropy is related to the boundary between regions I and II. Furthermore, one can see that $K_{\perp,1}$ increases with the Pt thickness up to $t_{Pt}$ = 2 nm, whereas the value of $K_{\perp,2}$ is growing just up to 1 nm. Above this thickness, $K_{\perp ,2}$ is rather stable, and its values are more or less 0.2 $MJ/m^3$. The $K_{\perp,1}$ reaches its the highest value at $t_{Pt} \approx 2 nm$ when it drops to the level about 0.75 $MJ/m^3$. The different values of $K_{\perp,1}$ than $K_{\perp,2}$ we relate to the much more efficient perpendicular anisotropy at Co/MgO than Co/Ti interface. On the other hand, the mean value magnetization saturation averaged over the whole range of the Pt thickness is about 1.07 T for both Co layers. The actual value for a given $t_{Pt}$ differs by $\pm$ 10 $\%$. The abrupt decrease in $M_S$ for very thin Pt layer (0.09 nm) is caused by the quality of the interfaces and related inter-mixing of Pt and Co atoms. 

The last but not least parameter derived from the macrospin is the interlayer coupling energy $J$. The polarization of the Pt is the mechanism of the interaction between two magnetic moments in Co layers. Such an interaction is ferromagnetic by its very nature,\cite{omelczenko} whereas the dipolar coupling (neglected here) is antiferromagnetic. The indirect way to probe the coupling (and the polarization of the Pt) by electrical detection is to measure FMR by rectification of radiofrequency current.\cite{harder,zietekPRB15,zietekAPL15} We derived the coupling ($J$) from the macrospin simulations, similarly as magnetizations and anisotropies. The coupling dependence on Pt thickness is shown in Figure \ref{fig:parametry}d. However, we could estimate $J$ down to Pt thickness of 1.36 nm, at which $J$ = 5.5 $mJ/m^2$. Below this thickness, the coupling has no effect on the resonance fields at frequencies experimentally accessible (< 25 GHz). The strong coupling causes both Co layers to rotate in the same manner, and they can be treated as one layer rather than two separate layers. In addition, two magnetizations of Co layers oscillate in phase (acoustic mode). It is seen in the experiment as observed low-frequency mode.\cite{ogrodnikPRM} On the contrary, the magnetization oscillations with opposite phases correspond to high-frequency optical mode (> 30 GHz), not achievable in the experimental method due to large losses in the power of the microwave current injected to the sample.\cite{zietekAPL15}
For this reason, although the exact value of $J$ is undeterminable for $t_{Pt} < 1.36 nm$ we just kept the value $J=5.5 mJ/m^2$ for simulations. This $J$ magnitude is marked as a horizontal dashed line in Figure \ref{fig:parametry}d. Its real value may reach any point from the hatched region, and particularly it may follow the exponential dependence, as predicted in Ref.\citenum{omelczenko} and shown in Figure \ref{fig:parametry}d too. For the thicker Pt ($t_{Pt} \ge 1.36 nm$, the fitting procedure to the Pt polarization model returned the Pt polarization depth parameter $\xi \approx 0.47 nm$, that is 1.5 times greater than $\xi$ reported for Py/Pt/Py structure.\cite{omelczenko}

Summarizing, we emphasize that the coupling strength correlates with the regions from I to IV. The constant value of $J$ within the region I corresponds to large and undetectable coupling, whereas in region II, $J$ is still significant and measurable. The intermediate region III is characterized by weak coupling, while samples within the region IV are  practically decoupled.

Having the magnetic parameters derived from the macrospin simulations of the spin-diode FMR dynamics, we calculated longitudinal static magnetoresistance ($R_{xx}$) dependencies on the external magnetic field in $H_x, H_y$, and $H_z$ directions. We also modeled the anomalous Hall resistance ($R_{xy}$) when the external magnetic field is applied in $z$ direction.

\subsection{Magnetoresistance and anomalous Hall effect \label{sec:MR}}
The Pt-based magnetic multilayers are expected to exhibit a large spin magentoresistance due to substantial spin-orbit interaction within the HM layer. These interactions cause the relatively large spin currents to be generated and injected into ferromagnetic
layers. The spin currents and spin accumulations at the Co/Pt interfaces influence the magnetoresistance of the sample, as predicted by the theoretical model presented in Sec.\ref{sec:lukasz}. Here we focus on the spin orbit interactions that are reflected in SMR. The SMR is defined as difference in longitudinal resistance measured in saturated magnetization of Co layers under the external magnetic field applied in $y$ and $z$ directions, i.e. $SMR = R_{xx}(H_y) - R_{xx} (H_z)$, while the AMR is defined similarly as in Sec.\ref{sec:lukasz} as $AMR = R_{xx}(H_x) - R_{xx} (H_z)$.\cite{kim} Also, we measured the AHE configuration ($R_{xy}$) in the field applied in $z$ direction. All magnetoresistance and AHE measurements were perfomed by DC current method sweeping the external magnetic field up to 10 kOe. 

\begin{figure}[H]
\includegraphics[width=6.5cm]{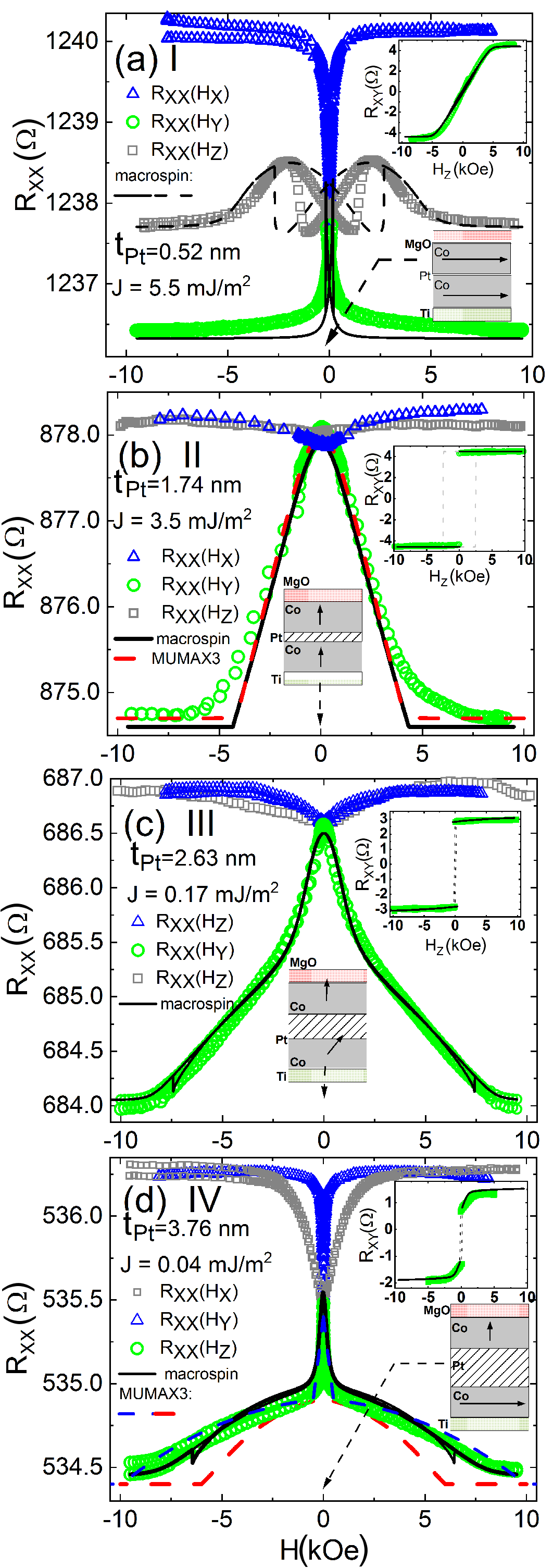}
\caption{The magnetoresistances $R_{xx}$ and $R_{xy}$ (inset)  as a function of the magnetic field in the samples with Pt thickness: (a) 0.52 nm as an example from region I, (b) 1.74 nm from region II, (c) 2.63 nm from region III and (d) 3.76 nm from region IV: $R_{xx}$ experimental data measured at magnetic field applied in $x$ (blue triangles), $y$ (green circles), $z$ (grey squares) directions. Depicted diagrams for all regions indicate the direction of magnetizations of magnetic layers at remanence. The macrospin (black solid and dashed lines) and micromagnetic(dashed red and blue lines) simulations of $R_{xx}(H_y)$. Micromagnetic simulations for $t_{Pt}$ = 3.76 nm were performed using the same parameters(cf. Figure \ref{fig:parametry}) as derived from macrospin model (red dashed line) as well as for an $K_1$ anisotropy increased by 0.17 $MJ/m^3$ (blue dashed line).}
\label{fig:figura2}
\end{figure}

Then, we modeled the magnetoresistance dependencies with use of the macrospin model. We used the parameters dervied previously by fitting the model to the FMR experimental results (shown in Figure \ref{fig:parametry}). For sake of simplicity, we treat the considered sample as a doubled bilayers: Co/Pt and Pt/Co. It allows us to calculate the resistance of Co/Pt/Co structure as the equivalent resistance of layers connected in parallel: $R_{xx,xy} = \frac{R_1 \cdot R_2}{R_1+R_2}$ where composite layer resistances are described by $R_{xx 1(2)} = R_{0,1(2)}+\Delta R_{AMR} m_{x 1(2)} ^2 + \Delta R_{SMR} m_{y 1(2)} ^2$,\cite{kim} while the AHE-related resistances are given by $R_{xy 1(2)} = \Delta R_{AHE} m_{z 1(2)} $.\cite{kim} In Figure \ref{fig:figura2} we show the typical MR curves for samples from regions I to IV. 

\begin{figure}[H]
\includegraphics[width=6.5cm]{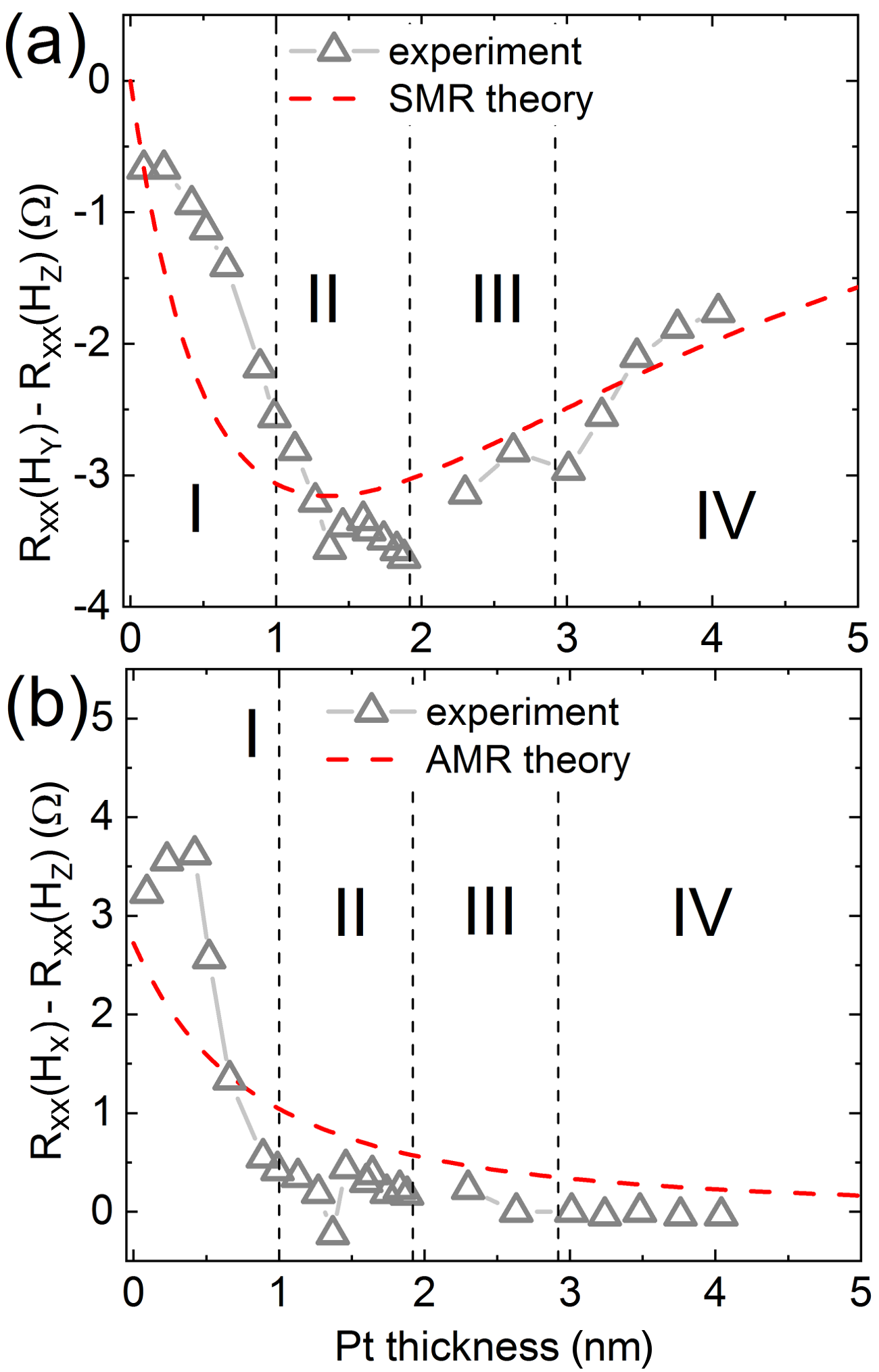}
\caption{SMR and AMR amplitudes derived directly from the measurements(empty grey triangles) and calculated within the diffusive model (red dashed lines) as function of the Pt thickness.}
\label{fig:figura3}
\end{figure}

For $t_{Pt}=0.52$ nm, the macrospin qualitatively reproduces a narrow peak in MR. It also accounts for a more complicated dependence of $R_{xx}(H_z)$ (see Figure \ref{fig:figura2}a). The AHE curve does not exhibit a hysteresis and its shape is typical for hard-axis rotation of both Co layers magnetized in-plane in the remanent state. The hysteresis in $R_{xx}(H_z)$ is due to a competition between a different anisotropies: in-plane and perpendicular that affect how the magnetizations rotate. Moreover, as one can see in Figure \ref{fig:figura3}b, the AMR effect dominates in the region I with the thinnest Pt layers. 

On the contrary, in region II, the SMR is the highest as predicted by the spin-diffusive model and macrospin simlutaions (cf. Figure \ref{fig:figura3}a). 
The representative sample from region II ($t_{Pt}$ = 1.74 nm) exhibits parabolic-like $R(H_y)$ dependence. It means that two Co layers are magnetized perpendicularly to the sample plane in the remanent state. Therefore the AHE reveals the clear switching-like shape (see inset in Figure \ref{fig:figura2}b). The both, simulation and experimental results, show the negligible contribution of the AMR.

For the sample from the region III ($t_{Pt} = 2.63 nm$), the $R(H_y)$ is rather convex-shaped than parabolic. On the other hand, the AHE curve still exhibits a switching-like behavior. It suggests that the magnetization of layer 2 is tilted away from the perpendicular towards in-plane direction. 

For the thickest Pt layer (e.g. $t_{pt}$ = 3.76 nm in Figure \ref{fig:figura2}d) when the Co layers are weakly coupled (region IV), one can see $R_{xx}(H_y)$ having parabolic-like shape in high magnetic fields. This part of the curve is due to rotation the perpendicularly magnetized Co layer from $z$ to $y$ direction. On the contrary, at low fields, there is a characteristic sharp peak related to rotation of the in-plane magnetized Co layer from its remanent state direction to the y direction. The dependence was well reproduced by macrospin model (black solid line in Figure \ref{fig:figura2}d). The same macrospin parameters provide the satisfactory agreement of AHE magnetoresistance with experimental points (see inset in the same figure). The AHE curve exhibits smooth-edged shape hysteresis, characteristic for the simulatneous rotation of the bottom Co layer magnetization ($\vec{M}_2$) in its hard-direction and switching of the top Co layer between two states: $\pm \vec{M}_{1,z}$.


We supported the macrospin model with micromagnetic simulations in the case of almost decoupled Co layers. The relevant calculations were performed with MUMAX3,\cite{Vansteenkiste2014} where the LLG equation was integrated numerically for each simulation cell. Due to memory and time usage limits, the simulated area was nominally restricted to 5 $\times$ 20 $\mu$m$^2$. However, we also utilized periodic boundary conditions along the $x$ direction in order to produce a demagnetization tensor matching the actual experimental conditions. To optimize simulation performance, the cell size was chosen as 4.88 $\times$ 4.88 $\times$ 0.87 nm for $t_{Pt}$ = 1.74 nm and as 4.88 $\times$ 4.88 $\times$ 0.94 nm for $t_{Pt}$ = 3.76 nm case. In both cases, the external magnetic field $H_y$ was increased with 500 Oe step, and the magnetization of Co layers was allowed to relax fully before moving to the next step. Then, the averaged magnetization vector for each layer was registered and used as an input for further resistance calculations. The micromagnetics revealed the same shapes of $R_{xx}$ curves as the macrospin model, for the same parameters (or very close), as shown in Figure \ref{fig:parametry} (see caption of Figure \ref{fig:figura2} for details). The agreement between macrospin and micromagnetic simulations confirmed that the macrospin parameters are reliable. 

For the sake of completness, in Figure \ref{fig:figura3} we show the $\Delta R_{AMR}$ and $\Delta R_{SMR}$ that were derived in whole range of the Pt thickness from experiment and prediced by the spin-diffusive model described in Sec.\ref{sec:lukasz}. The obtained amplitudes agree to a satisfactory extent. As one can see from Eqs. (\ref{eq:jchm}) and (\ref{eq:jcf}), the SMR and AMR depend on the charge current flowing in HM and F layers, respectively. However, the currents in HM layer are also influenced by spin accumulation at interfaces of this layer due to inverse SHE. The spin accumulation is mainly determined by the mean spin diffusion length ($\lambda_{HM}$) and spin Hall angle ($\theta_{SH}$). The negative value of the SMR reaches its maximum at $t_{Pt} \approx 1.5$ nm and decreases for thicker Pt layers, for which the spin decoherence affects the spin current, which, in turn, reduces the effective spin accumulation at F/HM interface. On the contrary, the positive value of the AMR rapidly and monotonically decreases with HM thickness since the average charge current density flowing into F layer decreases for thicker Pt layer.Discrepancies between the experimental and theoretical MR dependencies in the region I results from strongly mixed and alloyed interfaces for small thicknesses of Pt.

\subsection{Spin Hall angle and spin orbit torques \label{sec:SOT}}
\begin{figure}[H]
\includegraphics[width=8.5cm]{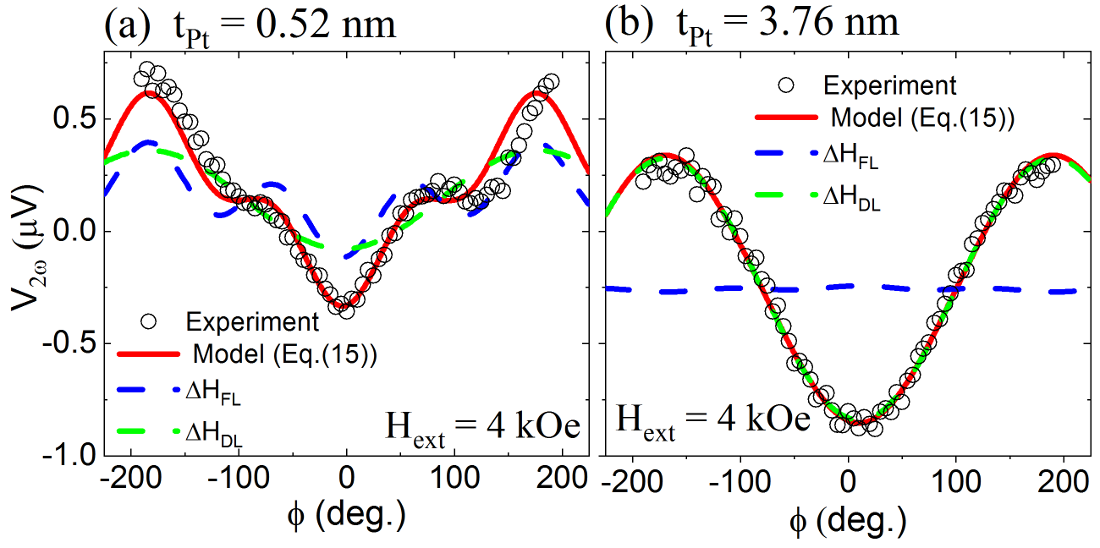}
\caption{Experimental harmonic voltages $V_{2\omega}$ for the representative samples from regions: (a) I ($t_{Pt}=0.52$ nm) and (b) IV ($t_{Pt}=3.76$ nm) both measured using the angular harmonic voltage method (cf. Eq.(\ref{eq:2nd})). The fitted trigonometric functions follow the Eq.(\ref{eq:2nd}).}
\label{fig:sotwyniki2}
\end{figure}

\begin{figure}[H]
\includegraphics[width=8.5cm]{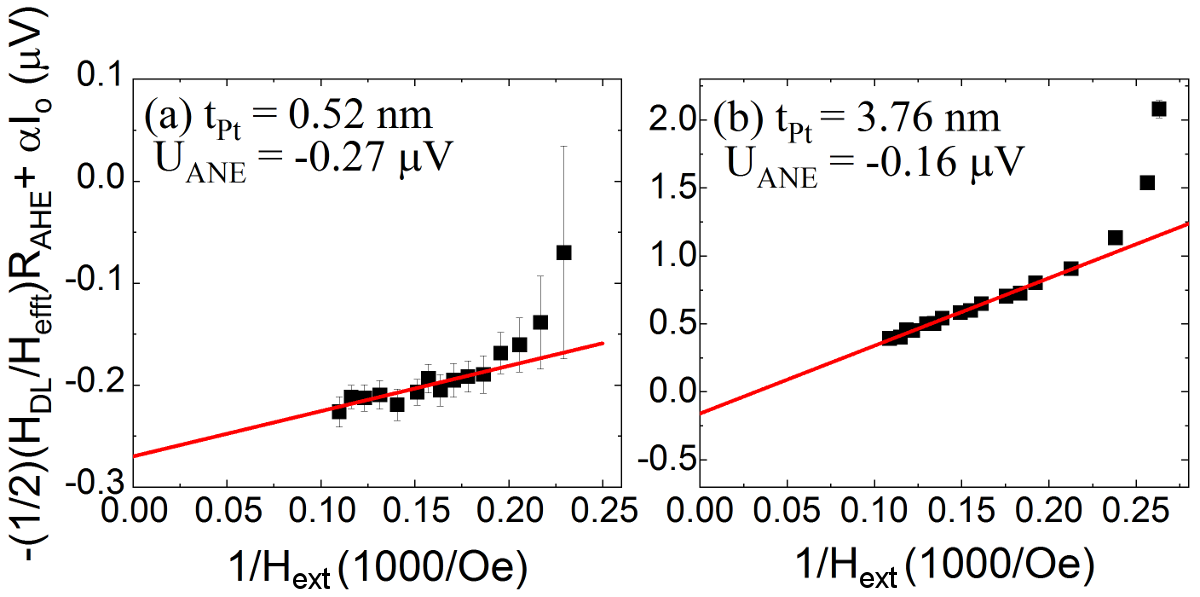}
\caption{The amplitude of Eq.(\ref{eq:2nd}) plotted as a function of magnetic field 1/$H_{ext}$ measured in the samples with Pt thickness 0.52 nm (a) and 3.76 nm (b) at angle $\phi_H = 45^\circ$. The interceptions with $y$ axis indicate the ANE contribution.}
\label{fig:sotwyniki3}
\end{figure}

\begin{figure}[H]
\includegraphics[width=8.5cm]{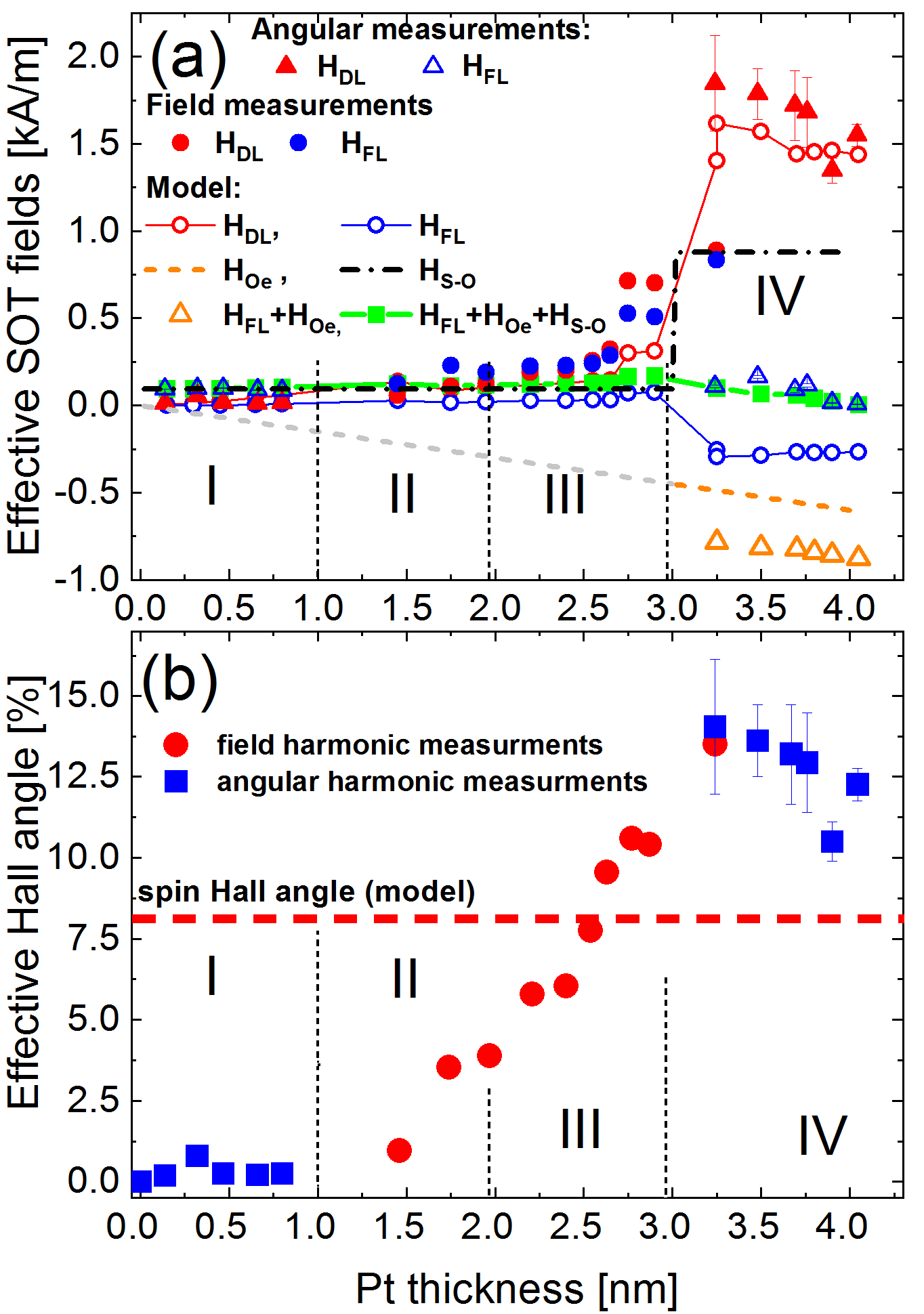}
\caption{ Experimental spin-orbit torque: damping-like and field-like components obtained with macrospin magnetic parameters from Figure \ref{fig:parametry} (blue and red filled points), the effective theoretical sum of SOT fields ($H_{FL}$ (blue empty points), and separately $H_{DL}$(red empty points)) acting on two Co layers. The amplitude of Oersted field($H_{Oe}$) (orange and grey dashed line) and Rashba-Edelstein spin-orbital field ($H_{S-O}$) (black dash-dotted line). The resultant field-like SOT (filled green squares) including $H_{S-O}$ field in region I and, additionaly, the Oersted field in region IV. Orange triangles correspond to the case when the Oersted field only is added to the $H_{FL}$. (b) The effective spin Hall angle determined from field (red circles) and angular (blue squares) harmonics measurements with the $H_{DL}$ amplitudes from (a). The red dashed line indicates the theoretical value of spin Hall angle fitted to the experimental MR dependencies (cf. Figure \ref{fig:figura3}) }
\label{fig:soty}
\end{figure}

In order to quantitatively characterize the spin-orbit interactions in the Co/Pt/Co trilayers, we performed the harmonic measurements briefly described in Sec.\ref{sec:harmonic}. 
For the samples for which both or one of the Co layers is magnetized in-plane (region I and IV) in the remanent state, we applied the angular harmonic voltage measurement method.\cite{avci2014,meinert} On the contrary, in the case of the Co layers that magnetizations are perpendicularly oriented (region II and III), we measured the field dependence of the relevant harmonic voltages.\cite{hayashi2014}
In the latter method, the damping-like (DL) and field-like (FL) components of SOT fields are determined using the following formula:
\begin{equation}
\Delta H_{DL(FL)} = - 2 \frac{B_{x(y)} \pm 2\xi B_{y(x)}}{1-4\xi^2}
\label{eq:torkisot}
\end{equation}
where $B_{x(y)} \equiv \frac{\partial V_{2\omega}}{\partial H_{x(y)}}/\frac{\partial^2 V_{\omega}}{\partial H_{x(y)}^2}$ and $H_{x(y)}$ stands for the in-plane external magnetic field applied in x(y) direction (cf. Figure \ref{fig:schemat}). 
The parametetr $\xi = \frac{\Delta R_{PHE}}{\Delta R_{AHE}}$ is the planar to anomalous Hall effect ratio.
\begin{figure}[h]
\includegraphics[width=8.5cm]{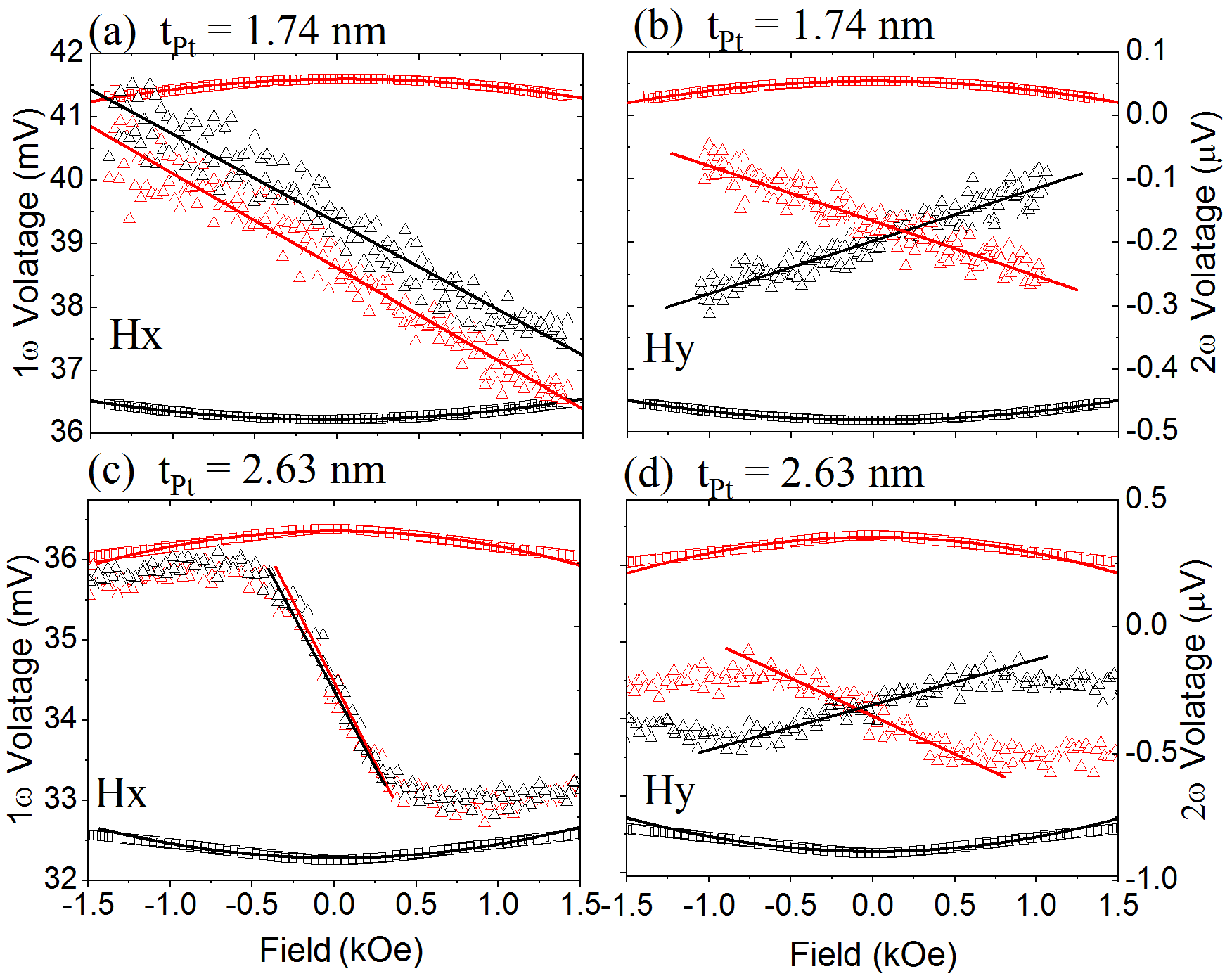}
\caption{Experimental harmonic voltages $V_{\omega}$ and  $V_{2\omega}$ for the representative samples from regions: (a-b) II ($t_{Pt}=1.74$ nm), (c-d) III  ( $t_{Pt}=2.63$ nm) both measured at the in-plane magnetic fields ($H_x$ and $H_y$) swept from -1.5 to +1.5 kOe (cf. Eq.\ref{eq:torkisot}). The fitted linear and quadric functions corresponding to the field-sweeping method (see Ref.\citenum{hayashi2014}). }
\label{fig:sotwyniki1}
\end{figure}
The first and second harmonic voltages ($V_{\omega}$ and $V_{2\omega}$) measured as a function of the applied magnetic field ($H_x$ and $H_y$) are plotted in Figure \ref{fig:sotwyniki1}. The results shown in the Figure \ref{fig:sotwyniki1} are representative for the samples from region II ($t_{Pt} = 1.74$ nm) and III ($t_{Pt} = 2.63$ nm).  Next, we could use the Eq.(\ref{eq:torkisot}) and follow the method described in Ref.\citenum{hayashi2014} to calculate SOT effective fields $\Delta H_{DL(FL)}$  in samples from regions II and III. The results are shown in Figure \ref{fig:soty}a and marked as squares. 
Nevertheless, the above method turns out to be ineffective in the case of samples with one or both layers magnetized in-plane. In this case, in order to determine $\Delta H_{DL(FL)}$, we measured the angular dependence of $V_{2\omega}$ on the magnetic field applied in the sample plane. Such a dependence may be expressed as follows\cite{meinert,avci2014} : 
$$V_{2\omega} = \left( -\frac{\Delta H_{FL}}{H_{ext}}R_{PHE} \cos 2\phi_H -\right.$$
\begin{equation}
\label{eq:2nd}
\left. - \frac{1}{2} \frac{\Delta H_{DL}}{H_{eff}}R_{AHE}   + \alpha_0  \right) I \cos\phi_H
\end{equation}
where $\phi_H$ stands for the in-plane angle of magnetic field. The term $\alpha_0$ is the anomalous Nernst effect(ANE) coefficient due to thermal gradients within the samples induced by the Joule heating.\cite{avci2014} The experimental angular dependencies $V_{2\omega}(\phi_H)$ for the samples with $t_{Pt}=0.52$ nm (region I) and $t_{Pt}=3.76$ nm (region IV) are shown in Figure \ref{fig:sotwyniki2}a,b. As one can see the damping-like SOT effective field ($\Delta H_ {DL}$) is proportional to the $\cos \phi_H$, whereas the field-like one ($\Delta H_{FL} $) is proportional to the $\cos\phi_H \cos 2\phi_H$.  Moreover, the $H_{eff} = H_{ext} - \frac{2 K_{eff}}{M_S}$, where $K_{eff}$(defined as in Sec.\ref{sec:FMR}) and $M_S$ are the parameters of the Co layers magnetized in-plane. As long as we knew the magnetic parameters (summarized in Figure \ref{fig:parametry}) of the layers, we could fit  Eq.(\ref{eq:2nd}) to the experimental data and consequenlty determine the both, field-like and damping-like SOT components.  
In addition, by plotting the terms proportional to $\Delta H_{DL}$ as a function of $1/H_{ext}$ we could estimate the contribution of the ANE. One should note that the offset of linear fit (at $1/H_{ext} = 0$) visible in Figure \ref{fig:sotwyniki3}a,b is the ANE contribution $\alpha_0 I_0$. We show the relevant plots for two samples ($t_{Pt}$ = 0.52 and 3.76 nm). The small offsets of fitted lines, ($\approx -0.3 \mu V$) and ($\approx -0.15 \mu V$) corresponding to the ANE-related electric fields $E_{ANE} \approx -0.03 V/m$ and $E_{ANE} \approx -0.015 V/m$ respectively, are much smaller than the values in the Co/Pt systems present in the literature, e.g. in Ref.\citenum{avci2014}. Therefore, it suggests that the ANE contribution is negligible in our devices with thin and thick Pt layers. 
It is worthy to notice that the dependencies shown in Figure \ref{fig:sotwyniki3} can be used to examine the applicability of Eq.\ref{eq:2nd}. At low magnetic fields the dependencies deviate from the linear and the Eq.\ref{eq:2nd} is not fulfilled. On the other hand, according to the model, the dependencies are linear at high fields. 

We summarize the results for the samples from all regions (I to IV) in Figure \ref{fig:soty}a. For very thin Pt (region I), both field-like and damping-like components are small, although the former contributes slightly more than the latter one. In the region II and III, both components increase in their magnitudes; however, in the region III, the DL component (filled red points), due to intermixing and alloying Co and Pt, surpasses the FL component's (blue filled points) magnitude. The most intriguing is region IV, where the damping-like component dominates over the field-like, especially for $t_{Pt}$ > 3 nm. For such a thick Pt layer, the $H_{DL}$ saturates, while the $H_{FL}$ drops again towards small values. Similarly the effective spin Hall angle defined by: $\theta_{SH,eff} = \frac{2e}{\hbar} \frac{H_{DL} M_S t_{Co} }{j_{Pt}}$ starts to increase in region II where the Pt thickness is sufficient to generate a significant spin currents, and consequently the $H_{DL}$ SOT field. The $\theta_{SH,eff}$ continues increasing in region III, and then it reaches its maximum value c.a. 14$\%$ at $t_{Pt} = 3. 24$ nm. Next it slightly decreases with the Pt thickness in region IV.

One should note that $\theta_{SH,eff}$ accounts for the spin accumulation effect at interfaces in the trilayer system. For this reason, the experimental value of $\theta_{SH,eff}$ may substantially differ from the theoretical one ($\theta_{SH}$ in Sec.\ref{sec:lukasz}), introduced as a material parameter of pure SHE efficiency at a very thick HM thickness limit). More precisely, the effective spin Hall angle is smaller than the theoretical one ($\theta_{SH}$) for $t_{Pt}$ < 2.6 nm. For thicker Pt layers, $\theta_{SH,eff}$ increases and becomes higher than the theoretical value. It is correlated with the change of magnetization direction from perpendicular to the in-plane for the thick Pt layer, particularly in region IV. In order to shed light on $\theta_{SH,eff}$ dependence on Pt thickness, one needs to go back to the Figure  \ref{fig:soty}a based on which the $\theta_{SH,eff}$ was determined. The two modes of experimental harmonics method allow the measurement of the effective SOT fields acting on the Co layers magnetized only in the plane (angular method) or perpendicular to the plane (field method). This limitation arises from very high fields required to saturate the layers magnetized in-plane (perpendicularly) at remanence into the perpendicular (in-plane) direction. Therefore, the data recorded with the experimental harmonic method accounts for both magnetic layers when magnetized in the same direction. That is the case of samples from regions I, II, and III. On the contrary, only one magnetic layer can be sensed by the measurement setup in region IV, where both magnetizations are orthogonal. Therefore, the experimental conditions differ in regions I-III and IV. As a consequence, the experimentally determined $H_{DL}$ and $H_{FL}$ fields in regions I-III are the sums of the related SOT fields acting on F1(top) and F2(bottom) magnetic layers. Since the spin currents generated by the SHE have opposite polarizations at F/HM and HM/F interface, the SOT effective fields also have the opposite signs. The residual fields, defined as $H_{DL}=H_{DL,1}+(-H_{DL,2})$ and $H_{FL}=-H_{FL1}+H_{FL2}$ come from the different interface properties, included in real parts of mixing conductances $G_{R1}$ and $G_{R2}$ of diffusive model(cf. Sec.\ref{sec:lukasz}). In order to support our analysis, we calculated the SHE-induced SOT effective fields using the formulas Eqs.(\ref{eq:HDL})-(\ref{eq:HFL}) and magnetic parameters from Figure \ref{fig:parametry}. Next, we plotted the difference of $H_{DL,1}$ and $H_{DL_2}$ in regions from I to III at the experimental thicknesses of Pt. Nevertheless, considering the experimental conditions discussed above, we plotted $H_{DL1}$ field only in region IV. The results are shown in Figure \ref{fig:soty}a as solid empty circles and agree with the experimental ones (red filled points - triangles and circles) to a satisfactory extent.

On the contrary, a similar procedure is insufficient to reproduce the experimental $H_{FL}$ field. The theoretical $H_{FL}$ (empty blue circles) depends differently on HM thickness than the experimental one (blue-filled circles and empty blue triangles). However, the diffusive model does not account for an Oersted field ($H_{Oe}$) coming from the charge current as well as a spin-orbital field ($H_{S-O}$) at interfaces, arising from the Rashba-Edelstein effect. Both fields have the same direction as SHE induced $H_{FL}$. Thus, the additional terms have to be included in the analysis of the effective field-like field. We assumed that the Oersted field linearly increases with the Pt thickness (grey and orange dashed line in  Figure \ref{fig:soty}a), whereas the $H_{S-O}$(black dash-dotted line) is independent of the HM thickness.\cite{staszek2019,witek2019} Since the Oersted field has the same amplitude with opposite signs in both F layers, its impact on the resultant $H_{FL}$ cancels out, and it does not affect the field-like SOT field in I-III regions. For the same reason, as discussed above, the $H_{Oe}$ adds to $H_{FL}$ in region IV (see full orange triangles in Figure \ref{fig:soty}a). Conversely, the $H_{S-O}$ fields do not cancel out due to the difference in F/HM and HM/F interfaces. Therefore their difference contributes to the total $H_{FL}$ term in regions I-III. In region IV the amplitude $H_{S-O}$ significantly increases because the perpendicularly magnetized Co layer is out of the experimental detection. Thus, the estimated magnitude generated at the Co/Pt interface was 0.86 kA/m. One should note that such a substantial value of the SOT field due to REE is much higher than its FL counterpart coming from the SHE. Moreover, it is of the order of the charge-current induced effective Oersted field. However, all fields SHE-related $H_{FL}$, $H_{Oe}$ and $H_{S-O}$ must be considered in the total FL SOT component in order to achieve a satisfactory agreement with experimental data in F/HM/F trilayer system (see solid blue line in Figure \ref{fig:soty}a). The discrepancies at the border between regions III and IV are due to the unreliability of the experimental methods (field and angular harmonics) applicable only when magnetic layers are magnetized fully in the plane or fully perpendicularly to the plane. This requirement is not fulfilled in the intermediate case, especially at $t_{Pt} = 2.87$ nm. It has its consequence in spin Hall angle value that slightly drops at this Pt thickness. On the other hand, at $t_{Pt} = 3.24$ nm, the SOT fields determined from the field and angular harmonics methods differ. In this case, the field harmonics method detects the magnetic layer magnetized perpendicularly. Therefore, as discussed earlier, the SOT fields at the Pt/Co interface are different from the SOT fields at the Co/Pt interface. However, the spin Hall angle (Figure \ref{fig:soty}b) treated as the HM material parameter is the same for both experimental methods. For thicker Pt, the detection signal in the field harmonics measurement was too weak to properly determine the $H_{DL}$ and $H_{FL}$ effective fields. 

\section{Conclusions\label{sec:suma}}
The paper presents the detailed results on structural, static, and dynamic properties of Co/Pt/Co trilayer in a wide range of the Pt layer thickness. First, we showed the Co/Pt and Pt/Co inherent interface asymmetry that resulted in different interfacial spin-orbit-related properties of ferromagnetic layers, magnetic anisotropies, and the effective spin-orbit torque fields due to SHE and REE. The difference in anisotropies makes the Co layer with stronger perpendicular anisotropy be a primary layer that determines the magnetization direction of the secondary Co layer through the interlayer exchange coupling. Therefore we were able to determine four ranges of Pt thickness where the trilayer reveals different static and dynamic behaviors correlated with the strength of coupling: region I (two Co magnetizations are in-plane), region II (Co magnetizations are both perpendicular to the plane), region III (one of Co magnetization is tilted from the perpendicular direction) and, finally, region IV (one Co magnetization is in-plane, whereas the second one is perpendicular). We showed that the experimental relation of dispersions and magnetoresistances differ in each region. This difference is accounted for by the macrospin models that we used, and therefore, both the experimental magnetoresistance and SOT-FMR relations of dispersion were reproduced by theoretical calculations to a satisfactory extent.
Moreover, we made a detailed analysis of the SOT effective fields determined using harmonics measurements. We showed that the experimental method applied to trilayers with two Co magnetizations aligned both in-plane or both out-of-plane allows measuring a difference of the effective SOT fields coming from two F/HM and HM/F interfaces. However, when two magnetizations are orthogonal, the experimental technique enables measuring SOT fields from the single F layer. The experimental results revealed this feature and were successfully parametrized with the magnitudes of damping-like and field-like SOT fields obtained from the diffusive model. Finally, both experimental and theoretical data allowed us to determine the contribution of Oersted ($H_{Oe}$) and spin-orbital ($H_{S-O}$) fields to the resultant experimental field-like SOT. We showed that the latter contribution due to REE might be comparable to the effective Oersted field and more significant than the field-like SOT field caused by the SHE.

\section*{ACKNOWLEDGMENTS\label{sec:acknow}}
This work was supported by the National Science Centre, Poland, Grant No. 2016/23/B/ST3/
 01430 (SPINORBITRONICS). 
Numerical calculations were supported by PL-GRID infrastructure. As part of cooperation, the multilayer systems were deposited in the Institute of Molecular Physics Polish Academy of Sciences and nanofabrication was performed at the Academic Centre for Materials and Nanotechnology of AGH University of Science and Technology. We would like to express our gratitude to Prof. F. Stobiecki for helpful discussions on data analysis. 
\bibliography{bibliography}
\end{document}